\documentclass[11pt, twocolumn, aip, apl,%
 showkeys,
 amsmath,amssymb, %preprint,%
 reprint
]{revtex4-1}

\usepackage[breaklinks=true,colorlinks=true,linkcolor=blue,urlcolor=blue,citecolor=blue]{hyperref}
\usepackage{graphicx}%
\usepackage{upgreek}%
\usepackage{dcolumn}%
\usepackage{soul,color}
\usepackage{bm}% bold math
\usepackage[mathlines]{lineno}% Enable numbering of text and display math
\expandafter\ifx\csname package@font\endcsname\relax\else
 \expandafter\expandafter
 \expandafter\usepackage
 \expandafter\expandafter
 \expandafter{\csname package@font\endcsname}%
\fi
\usepackage[utf8]{inputenc}
\usepackage[T1]{fontenc}
\usepackage{mathptmx}

\usepackage{siunitx} %User added package
\DeclareSIUnit{\belmilliwatt}{Bm}
\DeclareSIUnit{\dBm}{\deci\belmilliwatt}

\usepackage{miller} %User added package
\usepackage{xcolor} %User added package
\usepackage{ulem} %User added package

\usepackage{xfrac}
\usepackage{attachfile2}

\begin{document}
%TC:ignore
\preprint{AIP/123-QED}

\title{1D YIG hole-based magnonic nanocrystal} %
\date{\today}%
\author{K. O. Levchenko} 
\email{khrystyna.levchenko@univie.ac.at} % Corresponding author
\affiliation{Faculty of Physics, University of Vienna, Vienna, Austria}%
\author{K. Davídková}
\affiliation{Faculty of Physics, University of Vienna, Vienna, Austria}%
\affiliation{Vienna Doctoral School in Physics, University of Vienna, Vienna, Austria}
\author{R. O. Serha}
\affiliation{Faculty of Physics, University of Vienna, Vienna, Austria}%
\affiliation{Vienna Doctoral School in Physics, University of Vienna,  Vienna, Austria}
\author{M. Moalic}%
\affiliation{Department of Physics of Nanostructures, Adam Mickiewicz University, Poznań, Poland}%
\author{A. A. Voronov}
\affiliation{Faculty of Physics, University of Vienna, Vienna, Austria}%
\affiliation{Vienna Doctoral School in Physics, University of Vienna, Vienna, Austria}
\author{C. Dubs}
\affiliation{INNOVENT e. V. Technologieentwicklung, Jena, Germany}%
\author{O. Surzhenko}
\affiliation{INNOVENT e. V. Technologieentwicklung, Jena, Germany}%
\author{M. Lindner}
\affiliation{INNOVENT e. V. Technologieentwicklung, Jena, Germany}%
\author{J. Panda}
\affiliation{CEITEC BUT, Brno University of Technology, Brno, Czech Republic}
\author{Q. Wang}
\affiliation{Institute for Quantum Science and Engineering, HUST, Wuhan, China}%
\author{O. Wojewoda}
\affiliation{CEITEC BUT, Brno University of Technology, Brno, Czech Republic}%
\author{B. Heinz}%
\affiliation{Fachbereich Physik $ \&$ Landesforschungszentrum OPTIMAS, RPTU, Kaiserslautern, Germany}%
\author{M. Urbánek}
\affiliation{CEITEC BUT, Brno University of Technology, Brno, Czech Republic}%
\author{M. Krawczyk}%
\affiliation{Department of Physics of Nanostructures, Adam Mickiewicz University, Poznań, Poland}%
\author{A. V. Chumak}%
\affiliation{Faculty of Physics, University of Vienna, Vienna, Austria}%
\begin{abstract} 
Magnetic media with artificial periodic modulation—magnonic crystals (MCs) — enable tunable spin-wave dynamics and band structure engineering. Nanoscaling enhances these capabilities, making magnonic nanocrystals promising for both fundamental studies and applications. Here, we report on the design, fabrication, and characterization of one-dimensional YIG MCs with nanoholes ($d \approx 150$ nm) spaced $a \approx 1\,\upmu\mathrm{m}$ apart. Micro-focused Brillouin light scattering and propagating spin-wave spectroscopy, supported by TetraX and MuMax$^3$ simulations, reveal spin-wave transmission over 5 $\upmu\mathrm{m}$ in the Damon–Eshbach configuration, and the formation of pronounced band gaps with rejection levels up to 26 dB. Detailed analysis of the spin-wave dispersion uncovered complex mode interactions, including two prominent anticrossings at 3.1 and 18.7 rad/$\upmu\mathrm{m}$, between which the spin-wave energy is predominantly carried by the $n = 2$ mode, enabling efficient transmission. The results advance the development of functional MCs and open pathways toward 2D magnonic nanoarrays and magnonic RF nanodevices. %The combination of SWs advantages together with a flexibility of properties engineering through the means of patterning, makes magnonic crystals an excellent candidate for the RF applications, opening a path to the development of a single-mode transmission band device in GHz range, like magnonic RF stop-band filters \cite{16,17,74} and dynamic switching applications \cite{12,15,55}.

\end{abstract}%
\maketitle%

The field of magnonics explores the fundamental and applied potential of the spin waves (SW) - collective oscillations of magnetic moments in a magnetic material. The advantages offered by magnonics include high frequencies, tailored material parameters\cite{bottcher2022fast}, low energy dissipation and low power consumption, which have been successfully integrated into various prototype circuitry elements \cite{Chumak2014, talmelli2020reconfigurable}, with selected concepts surpassing the performance of benchmark conventional devices \cite{Wang.2020e}. Many of them (e.g., filters \cite{merbouche2021frequency}, transistors \cite{Chumak2014}, sensors \cite{inoue2011investigating, metaxas2015sensing}) are realized based on magnonic crystals (MCs) - artificial magnetic materials with a spatially periodic variation of properties \cite{Lenk2011, Krawczyk2014, Chumak2017}. Growing potential of nanoscale magnonics for RF applications, including magnonic crystals, was highlighted in a recent review \cite{levchenko2024review}. Similar to photonic crystals operating with light, MCs use the wave nature of their quasiparticles – magnons – to achieve propagation characteristics that are inaccessible by other means \cite{Puszkarski2003, Chumak2017}. Key properties of MCs, such as the central frequency and band-gap width, can be tailored by adjusting (often simultaneously) by (1) the use of different materials with suitable magnetic properties (e.g., saturation magnetization \cite{gubbiotti2010brillouin, obry2013micro, mruczkiewicz2017spin, dubs2025magnetically}), (2) the choice of the periodic pattern \cite{wang2013design} and (3) external factors, such as the applied magnetic field \cite{chumak2009current} or temperature \cite{vogel2015optically}. Different combinations of these properties produce a variety of MC designs, including waveguide- \cite{vogel2015optically, frey2020reflection} – and thin-film-based \cite{Puszkarski2003} structures; one-\cite{ikezawa2004preparation, chumak2009spin, klos2012effect, qin2018low,  roxburgh2024nano}, two-\cite{Tacchi.2010, tacchi2015universal} and three-\cite{Krawczyk2008, Krawczyk2010} dimensional; static \cite{obry2013micro, mruczkiewicz2017spin, wang2013design}, dynamic \cite{chumak2010all, chumak2009current}, and reconfigurable \cite{vogel2015optically, mantion2024reconfigurable} magnonic crystals.

\begin{figure*}
   \centering
   \includegraphics[width=\textwidth]{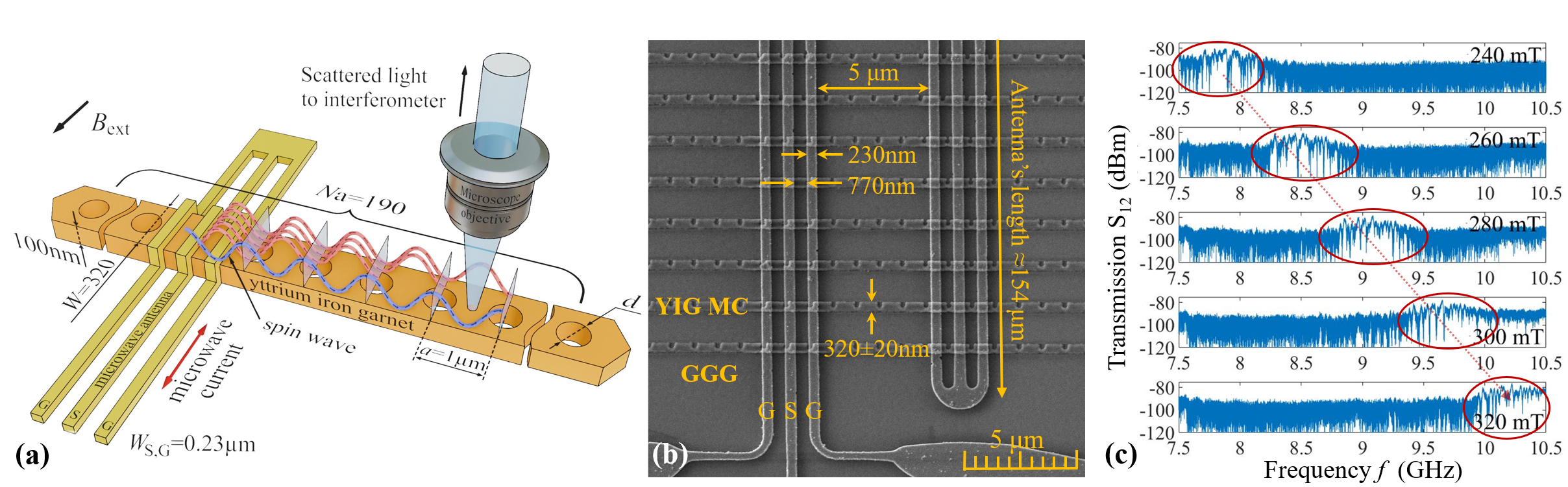}
   \caption{\label{Fig1}(a) Sketch of 1D waveguide-based MC periodically modulated with round holes. Spin waves excited by the CPW are shown in blue, while Bragg-reflected waves - in red. Key parameters: $\omega = 320$ nm – waveguide’s width; $\ a $ = 1 $\upmu$m – MC’s period; $\ N_a \approx 200$ – total number of periods; $\ d = 150 $ nm – diameter of the holes; (b) SEM image of the typical 1D MC fabricated from a 100 nm-thick LPE-grown YIG/GGG film, with key parameters as in (a). The number of waveguides per antenna is $\ n_\mathrm{wg}$ = 100, the distance between the antennas – 5 $\upmu\mathrm{m}$; (c) Spin wave transmission signal $\ S_{12}$ of magnonic crystal shown at (b) for varying bias magnetic fields in a frequency range 7.5 GHz - 10.5 GHz.}
 \end{figure*}

The simplest and most efficient type of magnonic crystal is a one-dimensional (1D) structure with geometric patterning. Typically, a 1D MC is fabricated from a waveguide (WG), where patterning enables the structure to act as a wavevector-dependent, mode-selective system \cite{Karenowska2010, Krawczyk2014}. 1D MCs can function as standalone devices or serve as building blocks for magnonic directional couplers, transistors, phase shifters, and other RF and data-processing components \cite{Chumak2014}. One of the fundamental works on this topic — a study on a geometrically patterned (notched) microscale permalloy 1D MC — was realized by Chumak et al. \cite{chumak2009spin}, based on a theoretical model by Lee et al. \cite{Lee2009} and numerical simulations by Ciubotaru et al. \cite{Ciubotaru2012}. The authors have experimentally demonstrated the propagation of coherently excited spin waves through a metallic magnonic crystal; however, in such macrostructured waveguides, the SW dispersion is inherently multimode \cite{demidov2015magnonic}, resulting in the simultaneous transmission of waves with different wavelengths at a fixed frequency. In magnonic crystals, this leads to band-gaps (BG) edges being less defined, forming a gradual slope in the transmission spectra \cite{chumak2009spin}, rendering their operating characteristics less favorable for applications \cite{Wang.2018}.

To overcome this limitation, MCs based on nanowaveguides can be considered. The effects of downscaling on the spin-wave spectra were explored by Wang et al. \cite{Wang.2019} and by Heinz et al. \cite{Heinz2020} When the width of an yttrium iron garnet (YIG) waveguide is sufficiently small, exchange interaction dominates over dipolar one, leading to unpinning of SW modes. This alters the quantization condition and shifts higher-order width modes to higher frequencies, effectively providing a single-mode regime \cite{Wang.2019}. Nanoscale enables key advantages for applications, as shown experimentally by Davidkova et al. \cite{davidkova2025nanoscale} in a multifunctional tunable magnonic nanodevice, or numerically by Ge et al.\cite{ge2024nanoscaled} in magnon nanotransistor. In the latter, notably, authors propose to use MC for precise control of the SW propagation and frequency-specific filtering. While nanoscale MCs hold great promise, their potential remains largely unrealized due to the relatively recent advances in nanofabrication.

Here, we report on experimental realization of a nanoscale 1D waveguide-based magnonic crystal, geometrically modulated with round holes. %To the best of our knowledge, such a structure has not been experimentally realized yet. 
Based on our preliminary studies, MCs with optimized geometrical parameters were designed and fabricated. Then, a detailed analysis of the spin-wave transmission in Damon-Eschbach (DE) configuration was performed with the means of Propagating Spin-Wave Spectroscopy (PSWS) across different frequency ranges, complemented by TetraX and MuMax$\mathrm{^{3}}$ simulations of the dispersion relation. Finally, microfocused Brillouin Light Scattering ($\upmu$-BLS) spectroscopy was carried out to demonstrate the spatial behavior of excitations in a single MC waveguide.

Figure~\ref{Fig1} presents a sketch of an individual MC WG with key parameters (a) and SEM image of a section of the entire fabricated structure (b). The crystal periodicity$\ a = 1~\upmu\mathrm{m}$ was selected to align with the maximum excitation efficiency of the antenna\cite{Vlaminck2010} (see Supplementary). The WG's width was designed as $w$ = 300 nm to support a single SW mode in a specific range, e.g., 8 - 8.2 GHz (<~5~rad/$\upmu\mathrm{m}$) under 262~mT bias field in DE and 10.3 - 10.9 GHz in Backward Volume configuration. After the fabrication, the width was around 320~nm, decreasing a single-mode window to around 100 MHz frequency bandwidth under same bias field. 
%3.1~rad/$\upmu\mathrm{m}$ < \textit{k} < 18.7~rad/$\upmu\mathrm{m}$ (8.2~GHz < \textit{f} < 9.16~GHz). 
The WG' length of $\approx 190~\upmu\mathrm{m}$ was chosen to ensure that no edge-reflected SW interfere with the measurements. The hole diameter$\ d$ modulates the SW reflection efficiency and thus affects the width and depth of the rejection bands \cite{Chumak2017}. Based on our simulations of similar structures, \textit{d} = 150 nm was estimated to provide the best ratio of minimized losses to well-defined BGs. The holes' pattern persisted along the whole waveguide. All structures were realized from LPE-grown 100 nm-thick YIG / GGG\,$(111)$ film \cite{Dubs.2017, Dubs.2020} using e-beam lithography and ion etching \cite{Heinz2020}. To estimate the spin-wave propagation length, multiple pairs of CPW antennas with varying spacings of 1, 2, 5 and 10 $\upmu\mathrm{m}$ were fabricated. Considering the small excited magnetic volume and potentially weak PSWS signal, each separate MC structure included up to 100 conduits. Coplanar waveguides (CPWs) were used for coherent spin-wave excitation and detection. Fabrication parameters matched the designed ones, apart from the conduits' width, as mentioned earlier, and slight upward shift of the holes.  %A typical mumax$^3$ simulation of the 1D MC in the Damon-Eschbach configuration (Magnetostatic surface spin wave) is shown in Fig.1(b), while results for the Backward and Forward volume configurations, as well as simulation parameters are provided in the Supplementary materials. The parameters used for the simulations are: $\omega$ = 300 nm; $\ a$ = 1 $\mu m$; $\ N_a$ = 200; antenna's wdith = 230 nm, and a modulation parameter $\ d_x$ varying in the range 5-250 nm, as indicated by color. The magnetic parameters of the 100 nm-thick YIG film were set to: $\mu_0 M_\mathrm{S} = 173 \pm2$ mT, in-plane anisotropy $\mu_0 (H_\mathrm{u} + H_\mathrm{c} ) = {5 \pm1}$ mT;  $\lambda_{\mathrm{ex}} = {3.8 \cdot 10^{-11} \pm0.15}$ Tm$^2$; $\alpha = {2\cdot 10^{-4} \pm0.2}% 

\begin{figure*}
    \centering
    \includegraphics[width=\textwidth]{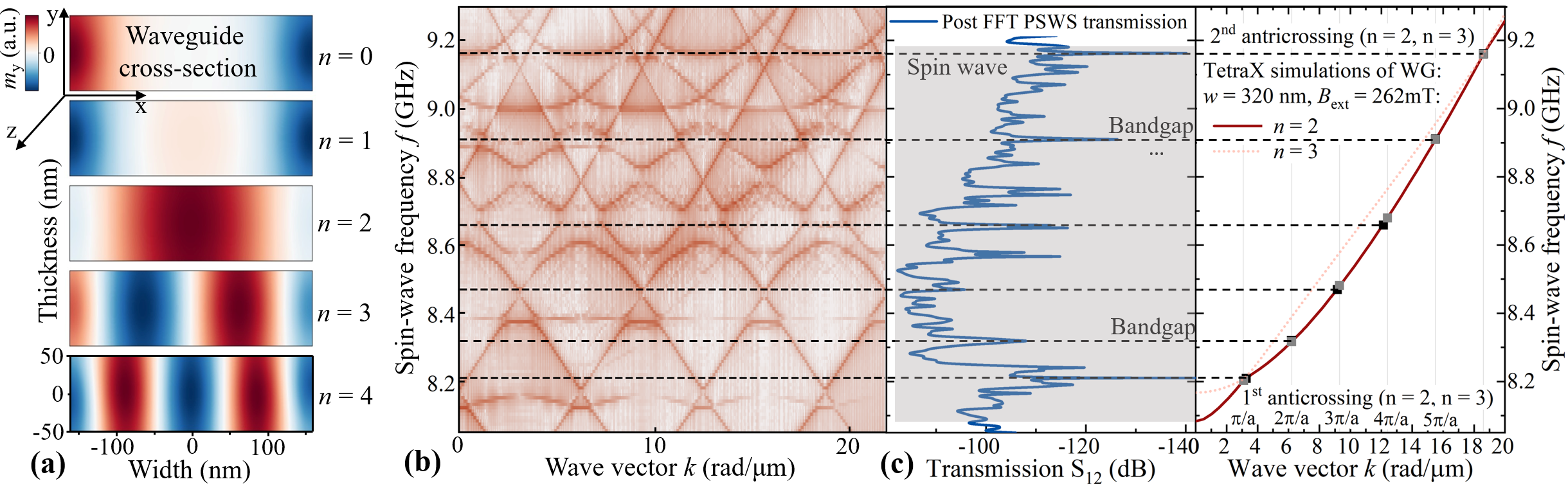}
\caption{\label{Fig2}(a) TetraX simulation of the mode profile amplitudes for the first 5 spin-wave modes of the unstructured waveguide in DE geometry with parameters as shown in Fig.~\ref{Fig1}(b). Real part of the magnetization dynamic component $m_\mathrm{y}$ is calculated for the waveguide cross-section at $k = 0$.  (b) MuMax$^3$ micromagnetic simulation of the 1D MC dispersion relation. (c) SW transmission $\ S_{12}$ at bias field $\approx 262 $ mT after 'time gating' post-analysis vs TetraX-simulated dispersion of a uniform waveguide in DE configuration. Black squares indicate the intersection of the wavenumbers $\ k_{a} = \pm n_{\mathrm{mc}}\pi / a $ at the band-gap frequencies with the simulated dispersion, while grey squares - with the experimental data, as indicated by the grey (c) and black (b) dashed projection lines. }
\end{figure*}

An exemplary propagating spin-wave spectrum, shown in Fig.~\ref{Fig1}(c), was measured on the 1D MC structure presented in Fig.~\ref{Fig1}(b). The SW transmission $S_{12}$ was recorded while applying a fixed microwave signal with -10 dBm power to the structure under test in DE configuration (in-plane, $ k \perp B_\mathrm{ext}$). The DE geometry was selected due to its higher excitation efficiency (compared to the Backward Volume, see Supplementary materials), enabled by higher SW group velocity and stronger coupling to the in-plane antenna field. The measured complex propagating SW signal, recalculated to dB magnitude (red circles, motion direction indicated by arrow), was obtained while sweeping the magnetic field from 240 mT to 320 mT in 20 mT steps across 7.5~GHz - 10.5~GHz frequencies. Reference background was subtracted as shown in our earlier work \cite{davidkova2025nanoscale}; other frequency and power ranges are provided in the Supplementary materials. The obtained SW spectrum significantly differs from that of a plain film by displaying 'band gaps' or 'rejection bands' – regions where propagation is prohibited and the signal's magnitude drops due to the Bragg’s scattering ($ n_{\mathrm{mc}}\lambda$ = 2$a \cdot$sin$\theta$, $n_{\mathrm{mc}}$ is an integer) of SW from the periodic holes \cite{Lenk2011, Krawczyk2014, Chumak2017}. Only the spin waves with wavenumbers $k_{a} = \pm n_{\mathrm{mc}}\pi / a $ satisfy this condition. The BG frequency depends on the material parameters, MC spatial geometry and can be tuned by the applied magnetic field, while number of gaps is defined by the Fourier distribution of spatial modulators\cite{Chumak2017}. Here, each hole introduces a sharp magnetic contrast (step-like function), generating multiple periodic band gaps. %Considering the $a = 1 \mu$m MC's modulation period and the $5\mu$m distance between the antennas in the measured structure, five band gaps are expected within the SW transmission signal. 
Accordingly, in 'transmission' or 'propagation' bands, where the Bragg condition is not satisfied, SW energy is expected to propagate without interruption, aside from higher insertion losses compared to an unstructured waveguide \cite{Chumak2017}. Losses at the level of $\approx$\,-80..-85\,dB are rather expected due to the low volume of magnetic material in structurally-modulated nanowaveguides. For reference, in a similar experiment by Davidkova et al. \cite{davidkova2025nanoscale} on a 97 nm-thick unstructured YIG film, the PSWS signal displayed losses of $\approx$~-25...-35~dB at -10~dBm power level. These losses can be further reduced up to four times through optimization of antennas' SW excitation efficiency\cite{bruckner2025micromagnetic}. Therefore, having a considerable number of the conduits within one MC structure was crucial for successful signal detection, despite increasing the risk of structural imperfections affecting the transmission.

The results of TetraX\cite{korber2021finite, korber2022tetrax} and Amumax~\cite{amumax2023} micromagnetic simulations (a fork of MuMax$^3$ ~\cite{mumax_2014,Leliaert2018}) are presented at Fig.~\ref{Fig2}(a,c) and (b), respectively. The WG was modeled according to the geometry in Fig.~\ref{Fig1}(b) with following YIG parameters: saturation magnetization $M_\mathrm{s} = 139$~kA/m, exchange stiffness $A = 3.7$~pJ/m,  uniaxial magnetic anisotropy $K_{\text{u}} = 3.58$~J/m$^3$ ($\perp$~x-axis) and Gilbert damping $\alpha = 10^{-4}$. Bias field $B_\mathrm{ext} = 262$~mT was applied along the y-axis. Mode profiles in Fig.~\ref{Fig2}(a) were calculated for the unstructured WG cross-section at $k = 0$, considering only the first five ($\textit{n}=5$) modes due to the quadratic decrease in dynamic magnetization intensity with increasing \textit{n}. The real part of the magnetization component $m_\mathrm{y}$  in the DE configuration is color-coded with red and blue (more in Supplementary). The first two modes ($n = 0, 1$) are edge modes, where SWs propagate only along the WG edges. Among them, only the $n = 1$ mode could be excited due to its symmetric mode profile. However, it cannot be resolved in the experiment because of its low amplitude and group velocity. Due to lateral confinement, the profiles show only the width quantization components $k_\mathrm{y} = n\pi /w $.  %Slight non-uniformity of profiles observed in BV configuration, e.g. in 0 mode, arises from plotting a single magnetization component, while contributions from both dynamic components $m_\mathrm{x}$ and $m_\mathrm{x}$ are important. With increasing frequency (and wavenumber, respectively) the excitation efficiency drops. 
 Under direct antenna excitation, only even width modes $(n = 2, 4, ...)$ are efficiently excited, as odd ones $(n = 3, 5, ...)$ have no net dynamic magnetization averaged across the width. In practice, slight antenna nonuniformity can excite odd modes, albeit inefficiently. Notably, while the mode profiles appear unpinned at the top/bottom surfaces, pinning persists at the WG' edges, which introduces elastic scattering into higher-order width modes. This occurs because of the larger waveguide's width than required for complete surface pinning and single-mode operation in a wider frequency range\cite{Heinz2020, Wang.2019}. \looseness=-1

The dispersion relation of the MC is shown on Fig.~\ref{Fig2}(b). To obtain it, the magnetization's x-components were recorded as functions of position and time, followed by a discrete Fourier transform along both axes for each simulation cell. The absolute value of the resulting complex spectra was computed and summed over the z- and y-axes to yield the final map. Black dashed lines correspond to the intersection of the estimated resonance conditions for Bragg scattering $\ k_{a} = \pm n_{\mathrm{mc}}\pi / a $ with micromagnetic simulation and experimental spin-wave transmission  (left panel of Fig.~\ref{Fig2}(c)). To obtain it, 'raw' data from Fig.~\ref{Fig1}(c) were subjected to 'time-gating'\cite{10696959} post-processing (details in Supplementary materials), improving the signal-to-noise ratio and increasing BG rejection efficiency through the elimination of main spurious signals. The highest signal amplitude originates from SWs with lower wavenumbers, driven by higher group velocity and efficient CPW excitation. Exceptions occur at anticrossing points, where mode hybridization leads to spatial localization and standing wave formation, resulting in a transmission drop below 140~dB. Six distinct BGs were identified in a spectrum: five arising from the hole-induced gaps between the antennas ($k_{1} = 3.1 \  \mathrm{rad}/\upmu\mathrm{m}$, $n_{\mathrm{mc}} = 1,..,5$) and two anticrossings ($k \approx$ 3.1 rad/$\upmu\mathrm{m}$ and $k \approx$ 18.7 rad/$\upmu\mathrm{m}$), the first coinciding with the structural gap. These anticrossings were revealed with TetraX dispersion simulation (Fig.~\ref{Fig2}(c) - right panel) for the modes $n_{\mathrm{mc}} = 2$ and 3 of the unstructured waveguide (crimson solid and peach dotted lines respectively). Grey squares mark intersections of the BG wavevectors $k_{a}$ (thin solid gray lines) with the simulations, while black squares - with the experimental data (black dashed lines). The rejection efficiency of the band gaps is between 12.6 dB and 26.1 dB. The dispersion reveals that the investigated 1D MC operates predominantly in a single-mode within the frequencies 8.08~GHz - 8.17~GHz (< 3.1~rad/$\upmu\mathrm{m}$). Reducing the waveguide width to approximately 280~nm is expected to eliminate mode anticrossings, while further narrowing to 250~nm would expand the single-mode frequency range sevenfold to 7.91~GHz - 8.56~GHz (up to 12~rad/$\upmu\mathrm{m}$; see Supplementary). We can also assume that within the linear regime, majority of the SW energy is effectively carried by \textit{n} = 2 mode in the range 8.2~GHz~-~9.16~GHz (3.1~-~18.7~rad/$\upmu\mathrm{m}$), since the edge modes \textit{n} = 0, 1 and the odd mode \textit{n} = 3 are not excited effectively. Both the MuMax$^3$ and TetraX simulations of an individual WG showed good agreement with the experiment, considering the cumulative PSWS signal from 100 conduits. The minor misalignment, along with the appearance of multiple weakly-defined gaps, is associated with fabrication imperfections. Variations in hole positioning and etching parameters among WGs cause reflections with slightly different amplitudes, phases, and wavelengths, leading to multiple Bragg conditions and the superposition of rejection bands. \looseness=-2

\begin{figure}[h]
    \centering
    \includegraphics[width=0.95\columnwidth]{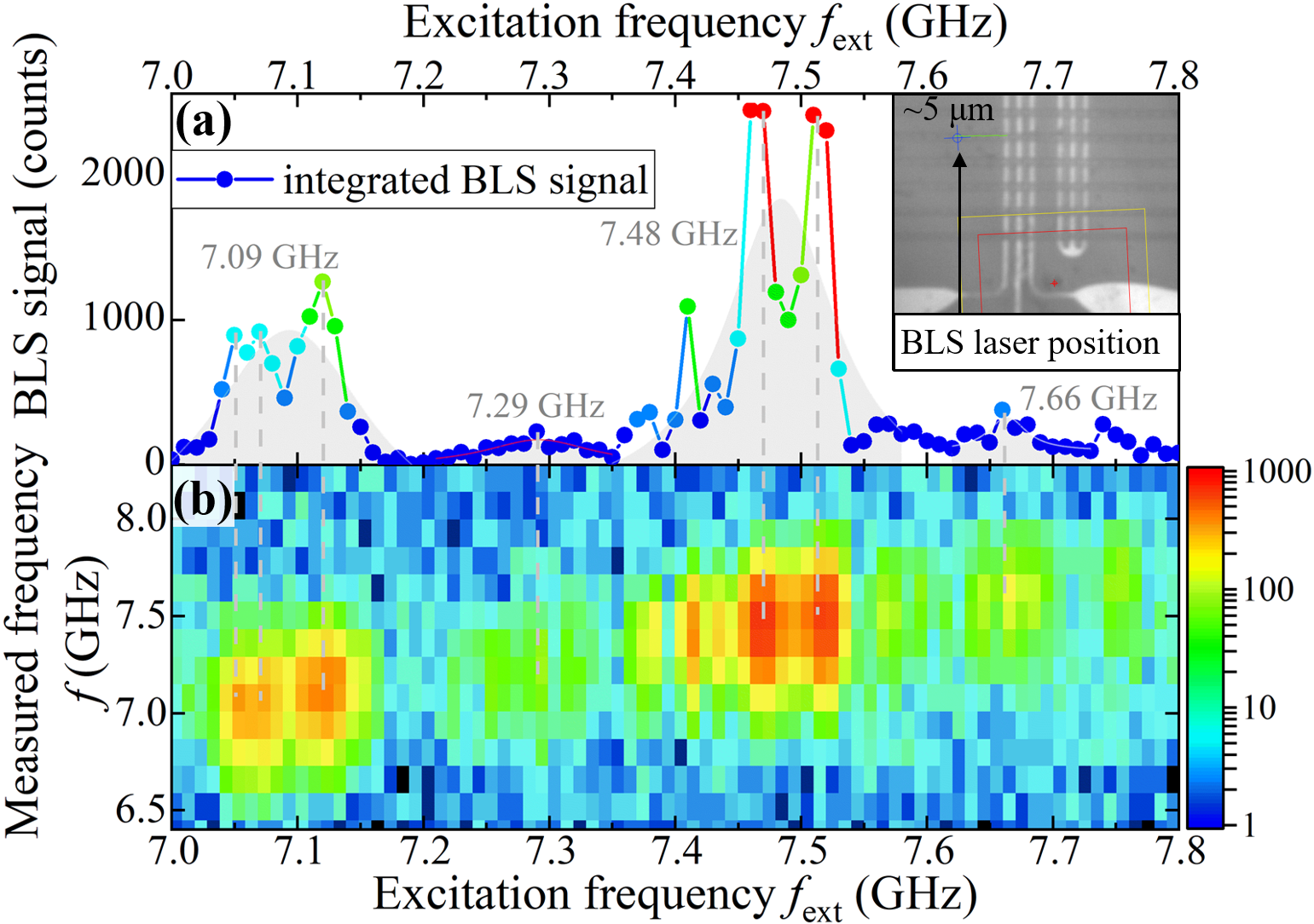}
    \caption{\label{Fig3}~BLS signal intensity of the propagating spin wave as a function of coherent excitation $ f_\mathrm{ext}$ (x-axis) in a form of: (a) 2D graph, where each y-axis point corresponds to the integrated BLS counts over respective excitation frequency $f_\mathrm{ext}$; (b) a 3D intensity map, where the y-axis shows the full range of measured frequency $f$ at each excitation frequency $f_\mathrm{ext}$, and BLS signal intensity (log scale) is color-coded at z-axis. Gray-shaded areas are fitted with a Lorentzian, with the peak frequency highlighted. Measurements were performed 5 $\upmu\mathrm{m}$ away from the coplanar waveguide’s antenna (inset). }
\end{figure}

After PSWS, coherently excited spin waves were probed by $\upmu$-BLS spectroscopy\cite{sebastian2015micro, wojewoda2024modeling}. An in-plane external field of $\mu_0 H=$ 238.6 mT was applied along the CPW antenna, ensuring uniform magnetization in DE configuration. Spectral analysis of the scattered light was performed using a 6-pass tandem Fabry-Pérot interferometer and a $\lambda_\mathrm{Laser}$ = 457 nm\cite{Hillebrands.1999} blue laser. Unlike PSWS, $\upmu$-BLS enables probing SW propagation within a single MC nanowaveguide. Figure~\ref{Fig3} shows the spin-wave signal, measured $\approx 5\ \upmu\mathrm{m}$ away from the CPW antenna (inset) on a MC with identical structural parameters to that analyzed by PSWS. Measurements were performed by sweeping the excitation frequency $f_\mathrm{ext}~=~7-7.8$ GHz in $\Delta f~=~0.01~$GHz steps, at a constant -10 dBm power. The SW signal in the form of BLS detector intensity (counts, Stokes part) is presented as a 2D map (Fig.~\ref{Fig3}(a)), where each y-axis point corresponds to the integrated BLS counts at respective $f_\mathrm{ext}$, and as a 3D intensity map (Fig.~\ref{Fig3}(b)), where the y-axis shows the full range of measured frequency $f$ at each excitation $f_\mathrm{ext}$; signal intensity color-coded as z-axis. BLS measurements were performed at the WG's center. While sweeping the microwave frequency, four periodically spaced passbands were observed (red and green areas in the 3D map), with the first signal appearing at 7.05~GHz and subsequent bands spaced by approximately 0.19~GHz. For clarity, adjacent peaks were Lorentzian-fitted (gray shaded areas), with the first peak, corresponding to the passband center, located at 7.09 GHz. Regions dominated by the background BLS counts (blue areas on the 3D map) represent five BGs measured 4.5 structural periods from the excitation antenna within a single MC waveguide. The diminishing BLS counts and the appearance of two-three separated peaks in the vicinity of 7.09 GHz and 7.48 GHz, indicate structural imperfections. %FROM HERE, PLEASE CHECK IF NOT BULLSHIT and may correspond to edge-localized modes in addition to propagating ones, and their possible hybridization. In patterned magnetic structures, inhomogeneous demagnetizing fields are formed at the edges of the holes, which can locally modify the internal magnetic field and effectively cause spin-wave confinement. As these modes are sensitive to structural parameters, such as hole size, spacing, and material thickness, fabrication imperfections of the thin films (which are more common due to the challenging nature of nanostructurization) contribute to the appearance of multiple peaks in the frequency spectra. Additionally, the generally lower signal intensity of propagating spin waves in thin films enhances the visibility of such parasitic modes. 

\begin{figure}[h]
    \centering
    \includegraphics[width=0.95\columnwidth]{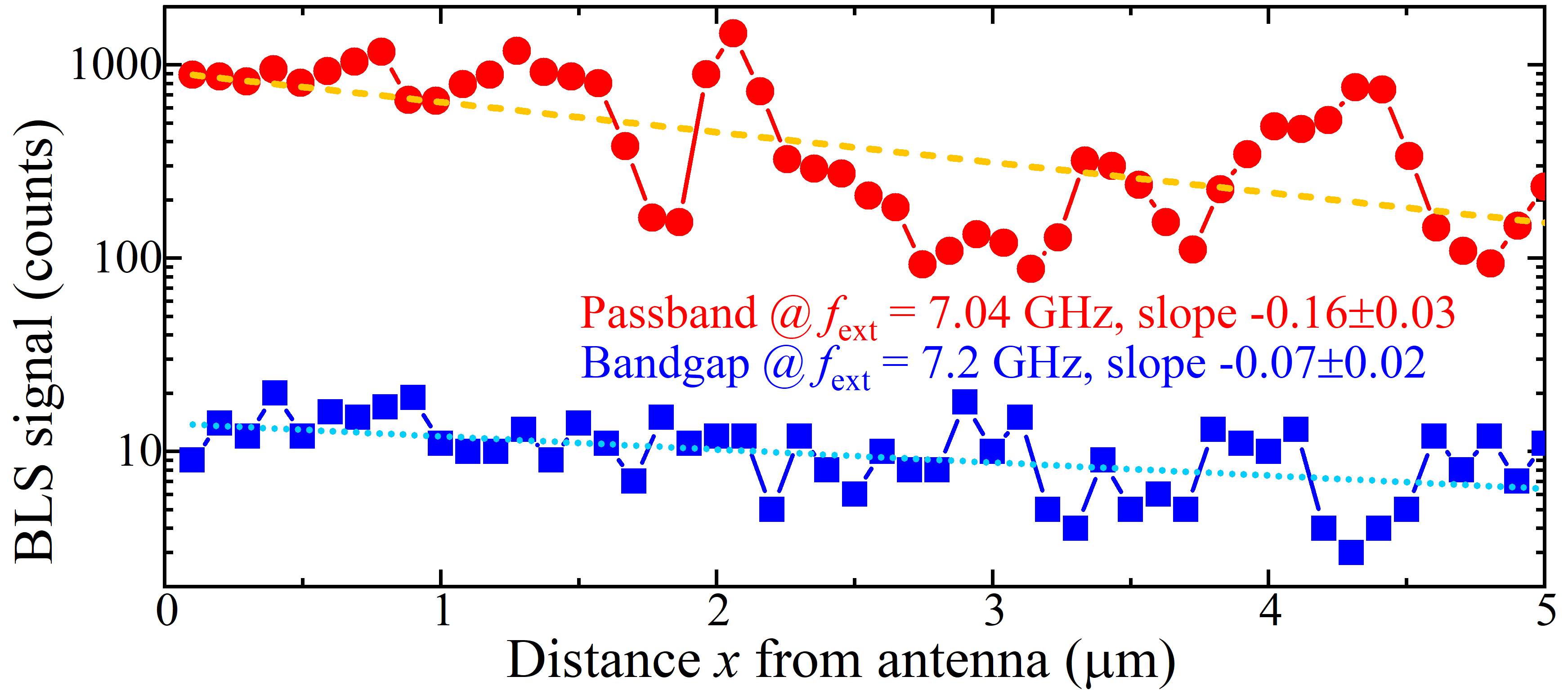}
    \caption{\label{Fig4} BLS signal intensity of the propagating spin wave as a function of a laser scan position 0-5 $\upmu\mathrm{m}$ from the antenna (x-axis); each y-axis point corresponds to maximum BLS counts at respective passband (red) or band-gap (blue) excitation frequency.}
\end{figure}

Finally, we investigated the spin-wave transmission in the passband and band-gap regions by sweeping the laser position from 0 to 5 $\upmu\mathrm{m}$ away from the antenna in $ \approx 100$ nm step. Figure~\ref{Fig4} demonstrates the maximum BLS intensity at the passband excitation frequency $ f_\mathrm{ext} = 7.04$ GHz (red circles) and at the band-gap frequency $ f_\mathrm{ext} = 7.2$ GHz (blue squares). The passband signal is an order of magnitude stronger than that of a bandgap, confirming the SW filtering. Linear fitting reveals a smaller slope for the bandgap signal (light-blue dots) compared to the passband (orange dashes), indicating suboptimal excitation efficiency and enhanced signal dissipation due to holes beneath the antenna. 

In conclusion, we demonstrated efficient spin-wave propagation in nanoscale one-dimensional YIG magnonic crystal modulated with holes. PSWS and BLS investigations revealed well-defined magnonic passbands and band gaps with a rejection efficiency up to 26 dB, corresponding to Bragg scattering from the periodic holes. Single-mode operation is achieved below the first anticrossing (<~3.1~rad/$\upmu$m) within a 100~MHz bandwidth, and can be further enhanced by narrowing the waveguides. Between the first and second anticrossings (1~GHz bandwidth, 3.1~-~~18.7~rad/$\upmu$m), most spin-wave energy is carried by the \textit{n}~=~2 mode, enabling effective SW transmission. While nanoscaling increases insertion losses and structural defects affecting the spectra, simulations confirm these to be only technical constraints. Future fabrication improvements are expected to firmly establish 1D YIG-based MCs as promising platforms for low-energy, high-frequency RF applications and magnonic computing.

\section*{Acknowledgements}
The research is funded by the Austrian Science Fund (FWF) project ESP 526-N TopMag [10.55776/ESP526] and by FWF IMEC [10.55776/PAT3864023]. MM and MK acknowledge Grant No. by National Science Centre of Poland (NCN) UMO–2020/37/B/ST3/03936 and 2023/49/N/ST3/03538. The work of M.L. was supported by the German Bundesministerium für Wirtschaft und Energie (BMWI) under Grant No. 49MF180119. B.H. acknowledges funding by the European Research Council within the Starting Grant No. 101042439 "CoSpiN". The authors thank Barbora Koraltan and Sabri Koraltan for the valuable discussions. 

\section*{Author contributions}
K.O.L. developed and coordinated the project, performed simulations, experiments, and analysis. K.D. conducted nanofabrication and provided essential support during the preliminary design, experimental stage, and PSWS analysis. M.M. and M.K. performed key micromagnetic simulations and assisted with data interpretation. R.O.S. significantly contributed to data analysis and experiments throughout the project. A.A.V. supported with all TetraX simulations. C.D., M.L. and O.S. grew and characterized the high-quality YIG nanofilms. J.P. and M.U. provided major support during the nanofabrication process. Q.W., O.W., and B.H. assisted in establishing the BLS experiments. A.V.C. conceptualized the article framework and contributed to important scientific discussions. All authors collaboratively refined and finalized the manuscript.
 
\section*{Competing interests}
The authors declare no competing interests.
%\section*{Additional information}

%\textbf{Reprints and permission information} is available at .
\section*{Data availability:}
The data that support the findings of this study are available
from the corresponding author upon reasonable request.

%\section*{Acknowledgments}
% This research was funded in whole or in part by the Austrian Science Fund (FWF) ESPRIT Fellowship Grant No. 10.55776/ESP526. 

%\section*{Data availability}
%The data that support the findings of this study are available from the corresponding author upon reasonable request.

%\nocite{}
\section*{References}
\bibliography{Bibliography}% Name of your .bib file

\clearpage
\onecolumngrid  % switch to one column

\clearpage
\thispagestyle{empty}
\noindent
\begin{picture}(0,735)
  \put(-31.5,0){
    \includegraphics[width=\paperwidth,height=\paperheight,keepaspectratio,page=1]{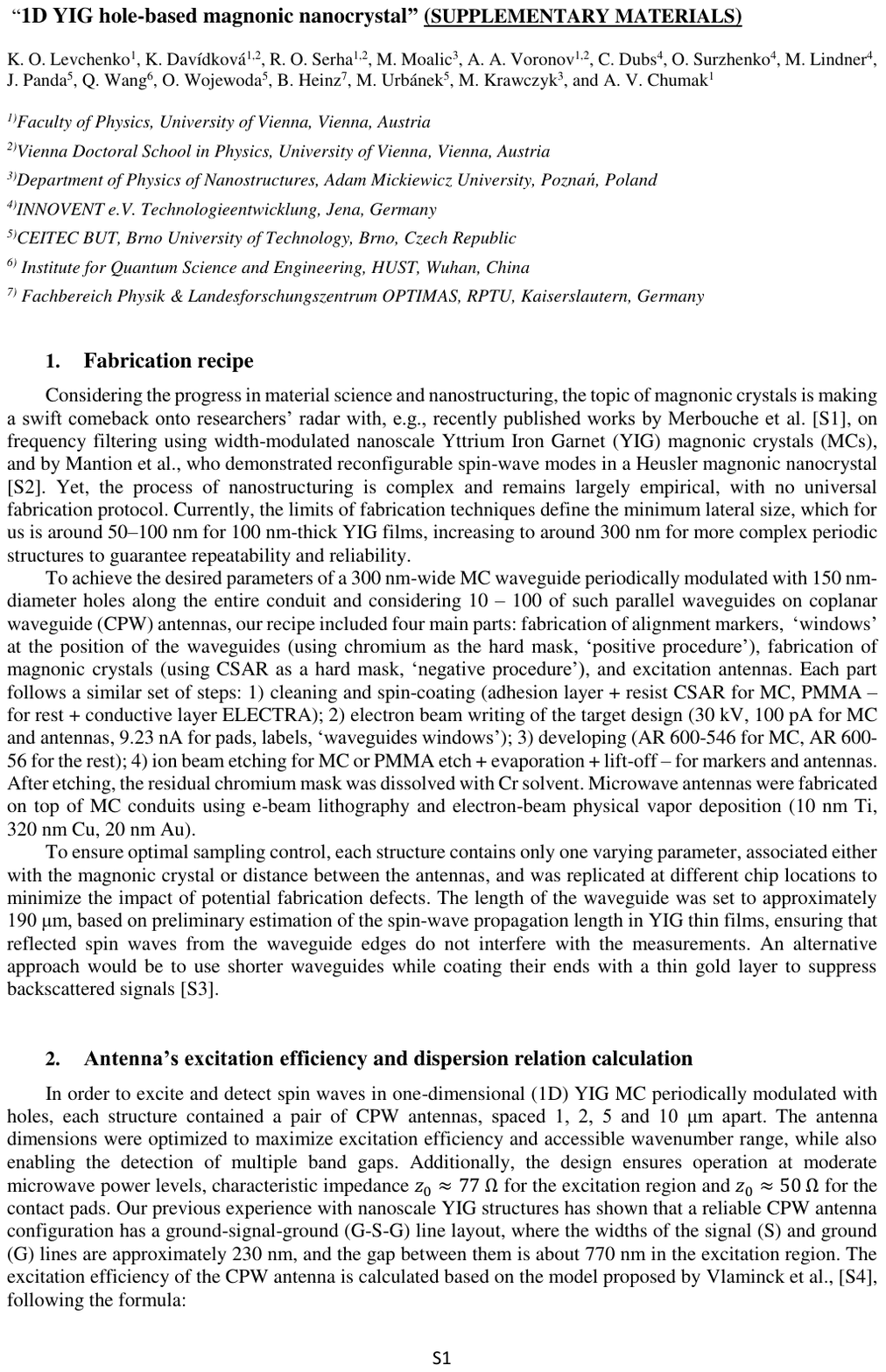}
  }
\end{picture}

\clearpage
\thispagestyle{empty}
\noindent
\begin{picture}(0,735)
  \put(-31.5,0){
    \includegraphics[width=\paperwidth,height=\paperheight,keepaspectratio,page=2]{SupMat.pdf}
  }
\end{picture}

\clearpage
\thispagestyle{empty}
\noindent
\begin{picture}(0,735)
  \put(-31.5,0){
    \includegraphics[width=\paperwidth,height=\paperheight,keepaspectratio,page=3]{SupMat.pdf}
  }
\end{picture}

\clearpage
\thispagestyle{empty}
\noindent
\begin{picture}(0,735)
  \put(-31.5,0){
    \includegraphics[width=\paperwidth,height=\paperheight,keepaspectratio,page=4]{SupMat.pdf}
  }
\end{picture}

\clearpage
\thispagestyle{empty}
\noindent
\begin{picture}(0,735)
  \put(-31.5,0){
    \includegraphics[width=\paperwidth,height=\paperheight,keepaspectratio,page=5]{SupMat.pdf}
  }
\end{picture}

\clearpage
\thispagestyle{empty}
\noindent
\begin{picture}(0,735)
  \put(-31.5,0){
    \includegraphics[width=\paperwidth,height=\paperheight,keepaspectratio,page=6]{SupMat.pdf}
  }
\end{picture}
\clearpage
\thispagestyle{empty}
\noindent
\begin{picture}(0,735)
  \put(-31.5,0){
    \includegraphics[width=\paperwidth,height=\paperheight,keepaspectratio,page=7]{SupMat.pdf}
  }
\end{picture}
\clearpage
\thispagestyle{empty}
\noindent
\begin{picture}(0,735)
  \put(-31.5,0){
    \includegraphics[width=\paperwidth,height=\paperheight,keepaspectratio,page=8]{SupMat.pdf}
  }
\end{picture}

\clearpage
\thispagestyle{empty}
\noindent
\begin{picture}(0,735)
  \put(-31.5,0){
    \includegraphics[width=\paperwidth,height=\paperheight,keepaspectratio,page=9]{SupMat.pdf}
  }
  
\end{picture}
\clearpage
\thispagestyle{empty}
\noindent
\begin{picture}(0,735)
  \put(-31.5,0){
    \includegraphics[width=\paperwidth,height=\paperheight,keepaspectratio,page=9]{SupMat.pdf}
  }
  
\end{picture}
\clearpage
\thispagestyle{empty}
\noindent
\begin{picture}(0,735)
  \put(-31.5,0){
    \includegraphics[width=\paperwidth,height=\paperheight,keepaspectratio,page=10]{SupMat.pdf}
  }
\end{picture}

\clearpage
\thispagestyle{empty}
\noindent
\begin{picture}(0,735)
  \put(-31.5,0){
    \includegraphics[width=\paperwidth,height=\paperheight,keepaspectratio,page=11]{SupMat.pdf}
  }
\end{picture}

\clearpage
\thispagestyle{empty}
\noindent
\begin{picture}(0,735)
  \put(-31.5,0){
    \includegraphics[width=\paperwidth,height=\paperheight,keepaspectratio,page=12]{SupMat.pdf}
  }
\end{picture}
\clearpage
\thispagestyle{empty}
\end{document}